\documentclass[sigconf]{acmart}

\AtBeginDocument{%
  \providecommand\BibTeX{{%
    \normalfont B\kern-0.5em{\scshape i\kern-0.25em b}\kern-0.8em\TeX}}}

\copyrightyear{2023} 
\acmYear{2023} 
\setcopyright{rightsretained} 
\acmConference[EASE '23]{Proceedings of the International Conference on Evaluation and Assessment in Software Engineering}{June 14--16, 2023}{Oulu, Finland}
\acmBooktitle{Proceedings of the International Conference on Evaluation and Assessment in Software Engineering (EASE '23), June 14--16, 2023, Oulu, Finland}\acmDOI{10.1145/3593434.3593470}
\acmISBN{979-8-4007-0044-6/23/06}

\begin{document}

\title{Identifying Characteristics of the Agile Development Process That Impact User Satisfaction}

\author{Minshun Yang}
\orcid{0009-0005-5281-1053}
\email{yangminshun@ruri.waseda.jp}
\affiliation{%
  \institution{Waseda University}
  \city{Tokyo}
  \country{Japan}
}
\authornotemark[1]

\author{Seiji Sato}
\orcid{0009-0008-7267-7738}
\email{r0d8h8i0h@asagi.waseda.jp}
\affiliation{%
  \institution{Waseda University}
  \city{Tokyo}
  \country{Japan}
}

\author{Hironori Washizaki}
\orcid{0000-0002-1417-9879}
\email{washizaki@waseda.jp}
\affiliation{%
  \institution{Waseda University}
  \city{Tokyo}
  \country{Japan}
}

\author{Yoshiaki Fukazawa}
\orcid{0000-0003-0196-2108}
\email{fukazawa@waseda.jp}
\affiliation{%
  \institution{Waseda University}
  \city{Tokyo}
  \country{Japan}
}

\author{Juichi Takahashi}
\orcid{0009-0005-6659-2015}
\email{juichi.takahashi@digitalhearts.com}
\affiliation{%
  \institution{AGEST Inc.}
  \city{Tokyo}
  \country{Japan}
  }


\begin{abstract}
The purpose of this study is to identify the characteristics of Agile development processes that impact user satisfaction. We used user reviews of OSS smartphone apps and various data from version control systems to examine the relationships, especially time-series correlations, between user satisfaction and development metrics that are expected to be related to user satisfaction. Although no metrics conclusively indicate an improved user satisfaction, motivation of the development team, the ability to set appropriate work units, the appropriateness of work rules, and the improvement of code maintainability should be considered as they are correlated with improved user satisfaction. In contrast, changes in the release frequency and workload are not correlated.
\end{abstract}




\received{10 March 2023}

\maketitle

\section{Introduction}

Improving user satisfaction is crucial not only in software development, but also for other product and service providers \cite{10.2307/3151722, CHI2009245, Kurdi2020TheIO, 9359262}. Although many empirical opinions about the processes that influence user satisfaction in Agile development exist, statistically supported research has yet to be reported to our knowledge.

This study aims to identify the characteristics of development processes that impact user satisfaction in Agile development. Specifically, the relationships between user satisfaction and development metrics, especially time-series correlations, are examined using data from user reviews and version control systems from OSS smartphone app developments. Some development metrics are correlated with user satisfaction. The results suggest that motivation of development teams, ability to set appropriate work units, appropriateness of work rules, and improvement in code maintainability are necessary but not exhaustive conditions to improve user satisfaction, whereas release frequency and workload are uncorrelated.

For researchers, this study suggested that a similar approach could be used to identify some more specific ominous smells in Agile development. For developers, specific ominous smells are presented and it will be possible to take care of them during development.

The rest of this paper is organized as follows. Section 2 reviews previous studies. Section 3 explains the methodology. Section 4 presents the results, which are discussed in Section 5. Finally, Section 6 concludes the paper and provides future prospects.

\section{Related Studies}

In a previous study of 623 Android apps, 77\% of apps that contained release notes similar to the important topics of their primary category had a statistically significant association with a positive change in the Google Play Store star rating \cite{8613795}. In contrast, similar studies, which analyzed code in Android apps did not find a relationship between code quality and user satisfaction \cite{10.1145/3194095.3194096}. Another study suggested that though reliability is a significant factor, once an acceptable level of reliability is achieved, three other factors dominate the user satisfaction: capability, usability, and performance by doing user interviews \cite{10.2307/2633041}.

However, these studies only focused on different aspects of development artifacts such as release notes but did not examine the development process. Herein the relationships between user satisfaction and the characteristics of the development process are examined. Additionally, time-series correlations are considered with and without a time delay. 

\section{Analyses of the Relationships Between Development Processes and User Satisfaction}

\subsection{Overview}

This study answered the following research questions:

\subsubsection{RQ1. Can the characteristics related to the Agile development of the development process be measured and evaluated?}\hfill

This study tries to identify development metrics \cite{KUPIAINEN2015143, 7112365, BIESIALSKA2021106448, LOPEZ2022111187} that are relevant to Agile development. Then the metrics are measured, and the characteristics related to Agile development processes are evaluated.

\subsubsection{RQ2. Is there a relationship between the characteristics related to the development process of Agile development and user satisfaction?}\hfill

\begin{figure}[tb]
  \centering
  \includegraphics[width=\linewidth]{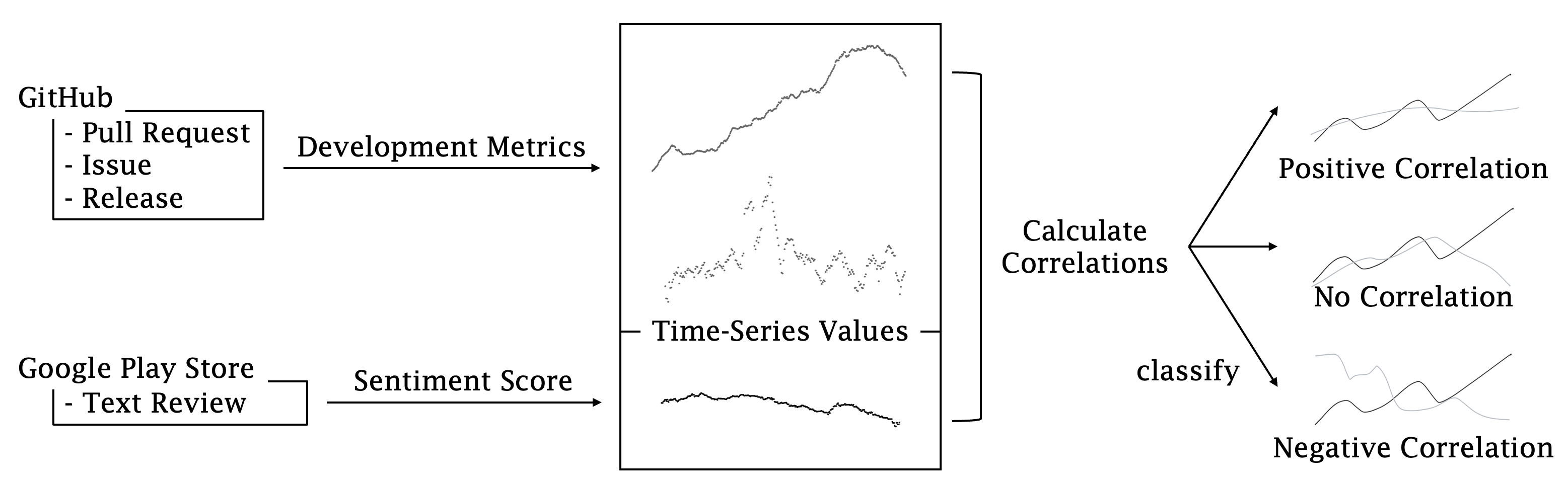}
  \caption{Overview}\label{fig1}
\end{figure}

This study analyzes the time-series correlations between the development metrics identified in RQ1 and the sentiment scores of user text reviews, which indicate user satisfaction. Figure \ref{fig1} overviews the process. Since the changes in the characteristics of the development process may not be immediately reflected in the changes in user satisfaction, a time delay was considered in the time-series correlations.

\subsubsection{RQ3. What approaches can be applied in the development process to improve user satisfaction?}\hfill

To answer this research question, possible approaches based on the time-series correlations revealed in RQ2 are analyzed using the factors affecting metric changes. Both significant and weak correlations are used in the analysis.

\subsection{User Satisfaction}

\begin{figure}[tb]
  \centering
  \includegraphics[width=\linewidth]{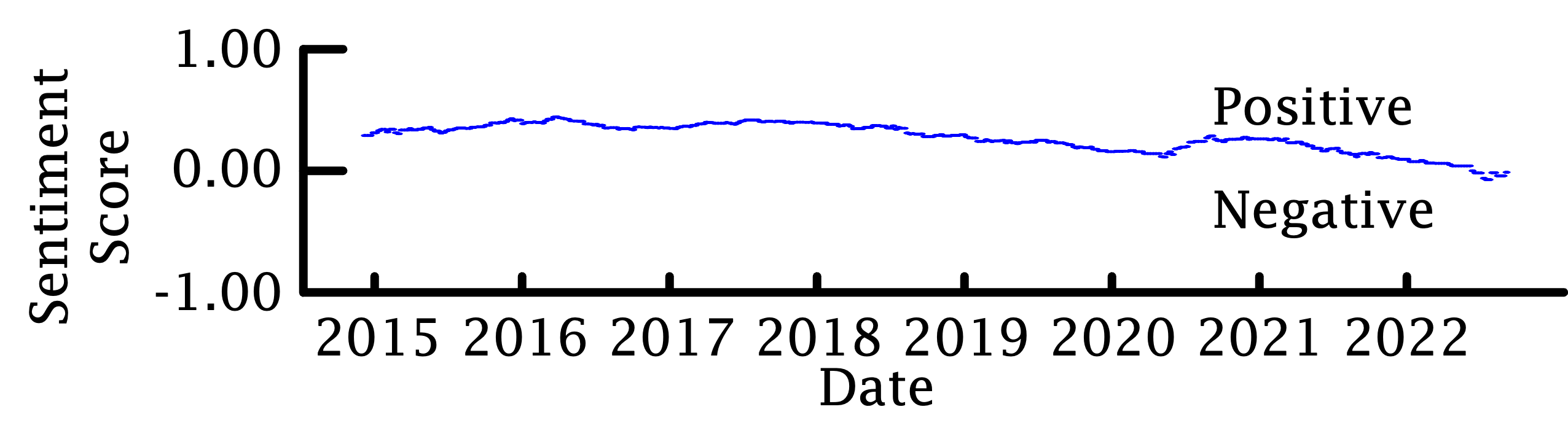}
  \caption{Example of a Time-Series Sentiment Score for Text Reviews}\label{fig2}
\end{figure}

\begin{table}
    \caption{Example of Sentiment Scores for Text Reviews}
  \label{tab:table1}
    \begin{tabular}{p{5.5cm}c}
    \toprule
        \multicolumn{1}{c}{Text Review} & Sentiment Score \\
    \midrule
        Just installed. Stuck at pull down to start. & –0.25 \\
        Beautiful to look at, but I am lost in UI. Swipe here,   swipe there I don't know where I am. & –0.128 \\
        Nice UI and features, very well use of pictures.   The App stands out from the rest, its attractive ! Waiting for Windows client   of Wire. & 0.8118 \\
    \bottomrule
    \end{tabular}
\end{table}

User satisfaction was indicated by the numerical value of the sentiment analysis applied to the user’s text review \cite{10.5555/3370272.3370278, 7332474, 8501305}. 

Here, VADER \cite{Hutto_Gilbert_2014} was used to estimate the sentiment of the inputted English sentences and outputted a numerical value from –1 (negative) to 1 (positive). This numeral value was subsequently used in the sentiment analysis. Table \ref{tab:table1} shows an example of the emotion analysis of a text review, while Figure \ref{fig2} gives an example of the output of the emotion analysis score as a time series.

\subsection{Characteristics of the Agile Development Process}

\begin{table*}
    \caption{Targeted Metrics on Processes}
  \label{tab:table2}
    \begin{tabular}{cccp{9cm}}
    \toprule
        Metric Name & No. & \begin{tabular}{c}Related 12 \\ Principles of Agile\end{tabular} & \multicolumn{1}{c}{Definition} \\
    \midrule
        Lead Time & M1 & 3, 7, 8 & Average time from issue submission to resolution and release within the aggregation unit. \\
        Number of Merged Pull Requests & M2 & 7, 8 & Total number of merged Pull Requests in the aggregation unit. \\
        Merge Duration & M3 & 7, 8 & Average time from the creation of a Review Request to its merge within the aggregation unit. \\
        Remaining Issues Lifetime & M4 & 7, 8 & Average survival time of issues in the Open state within the aggregation unit. Only issues that were finally closed by Pull Request are included. \\
        Remaining Pull Requests Lifetime & M5 & 7, 8 & Average survival time of Pull Requests in the Open state within the aggregation unit. Only Pull Requests that were finally merged are included.\\
    \bottomrule
    \end{tabular}
\end{table*}

The development metrics showed the characteristics of the Agile development process. We extracted metrics, which may influence user satisfaction and could be easily collected. The latter requirement was due to time constraints. In Agile development, the potential for a relationship with the changes in user satisfaction was judged by its relevance to the 12 principles of Agile \cite{HighsmithFowler2001}  in this study. Table \ref{tab:table2} shows the development metrics covered, while Table \ref{tab:table3} shows the number of the applicable Agile principle.

\begin{table}
    \caption{Number of the Applicable Agile Principles}
  \label{tab:table3}
    \begin{tabular}{cp{7cm}}
    \toprule
         No. & \multicolumn{1}{c}{Principle} \\
    \midrule
        3 & Deliver working software frequently, from a couple of weeks to a couple of months, with a preference to the shorter timescale.  \\
        7 & Working software is the primary measure of progress.  \\
        8 & Agile processes promote sustainable development. The sponsors, developers, and users should be able to maintain a constant pace indefinitely.  \\
    \bottomrule
    \end{tabular}
\end{table}

\subsection{Selection of Research Objects to Identify the Relationship between User Satisfaction and the Agile Development Process}

\begin{table}
    \caption{Selection Criteria for Research Objects}
  \label{tab:table4}
    \begin{tabular}{p{3cm}p{4.5cm}}
    \toprule
         \multicolumn{1}{c}{Purpose} & \multicolumn{1}{c}{Criteria} \\
    \midrule
        Available data & OSS \\
        No restrictions on obtaining user reviews & Available on Google Play Store \\
        Development metrics can be calculated & Used GitHub for development  \\
        Agile Orientation & More than 20 Contributors indicated by GitHub \\
        & Can confirm the use of CI tools \\
        & Several releases per year can be confirmed \\
        & Development can be confirmed using GitHub's Pull Request feature \\
        Eliminate short-term noise & There is more than 2.5 years between the first and most recent release of the app \\
        Sufficient data & Total number of GitHub Pull Requests is over 500 \\
        & Average of more than 40 text reviews written in English annually \\
    \bottomrule
    \end{tabular}
\end{table}

Table \ref{tab:table4} shows the selection criteria for the research objects used to identify the relationship between user satisfaction and the development process. Our research methodology required data that show the characteristics of user satisfaction and the development process. Due to data source limitations, the research subject was limited to OSS software because it is available to the public. Additionally, we further narrowed the scope to publicly available Android apps in the Google Play Store due to the accessibility of obtaining user reviews, which are an integral part of this study. The Apple Store has restrictions on obtaining user reviews, whereas the Google Play Store does not. Moreover, we targeted apps that used GitHub in their development since metrics indicating the characteristics of the development process can be calculated. Other requirements included a minimum number of 20 contributors indicated by GitHub, a record indicating several releases per year, and the development was conducted using the Pull Request feature. To remove short-term noise in the data analysis, the time between the first and most recent release must be at least 2.5 years. Finally, to ensure sufficient data, the total number of Pull Requests and annual average number of text reviews written in English must be at least 500 and 40, respectively.

The selection of research objects was based on a check of the OSS Android apps listed at \url{https://github.com/offa/android-foss#-apps-}. This list was selected due to its high update frequency, openness in terms of editing privileges, and the repository generally covers OSS Android apps.

\section{Results of Measurements and Analyses}

\subsection{Projects to Be Analyzed}

Based on the method described in Section 3.4, 37 apps in 22 different categories were selected from 374 apps. We assumed that the bias in selecting apps was negligible and did not affect the results.

\subsection{Measurement Results}

For the 37 selected apps, we calculated the time-series trends of each app’s user review sentiment score and the targeted development metric values using weeks as the aggregation unit. Then we calculated the time-series correlations between the sentiment scores of the user reviews and the development metrics for each app with delays ranging from no delay to 80 weeks. A correlation coefficient of 0.3 or higher indicated a positive correlation, while a value of –0.3 or lower denoted a negative one. All other values were assumed to be uncorrelated. If both positive and negative correlations were found, the trend with a stronger correlation coefficient was adopted.

\begin{figure}[tb]
  \centering
  \includegraphics[width=\linewidth]{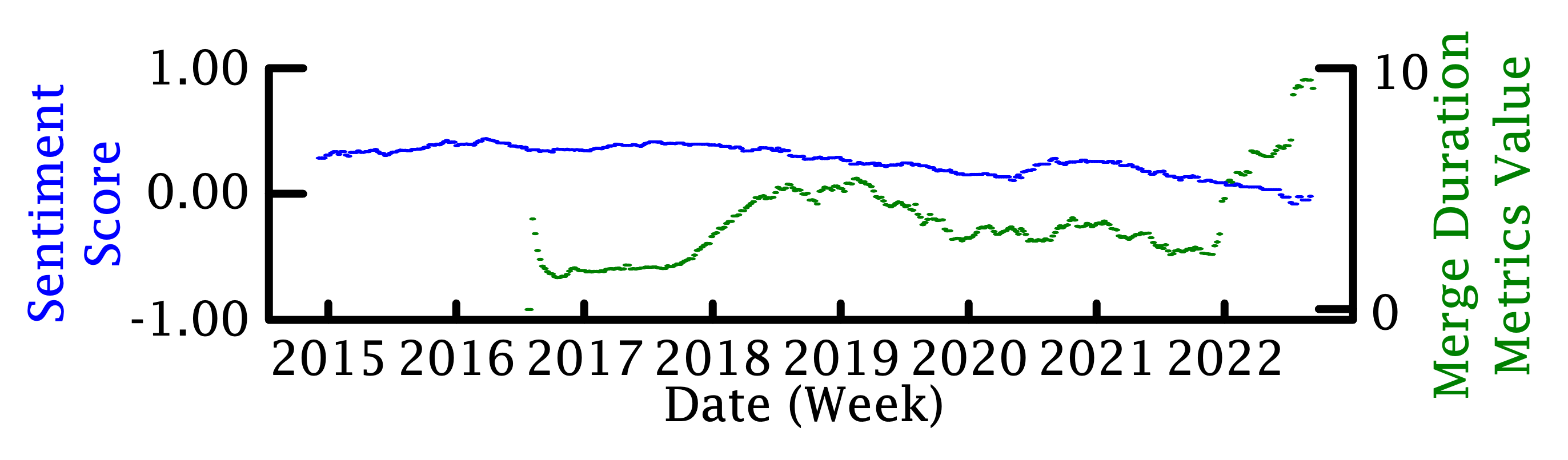}
  \caption{Time-Series Sentiment Score and Merge Duration Metrics Value of the Wire App}\label{fig3}
\end{figure}

\begin{figure}[tb]
  \centering
  \includegraphics[width=\linewidth]{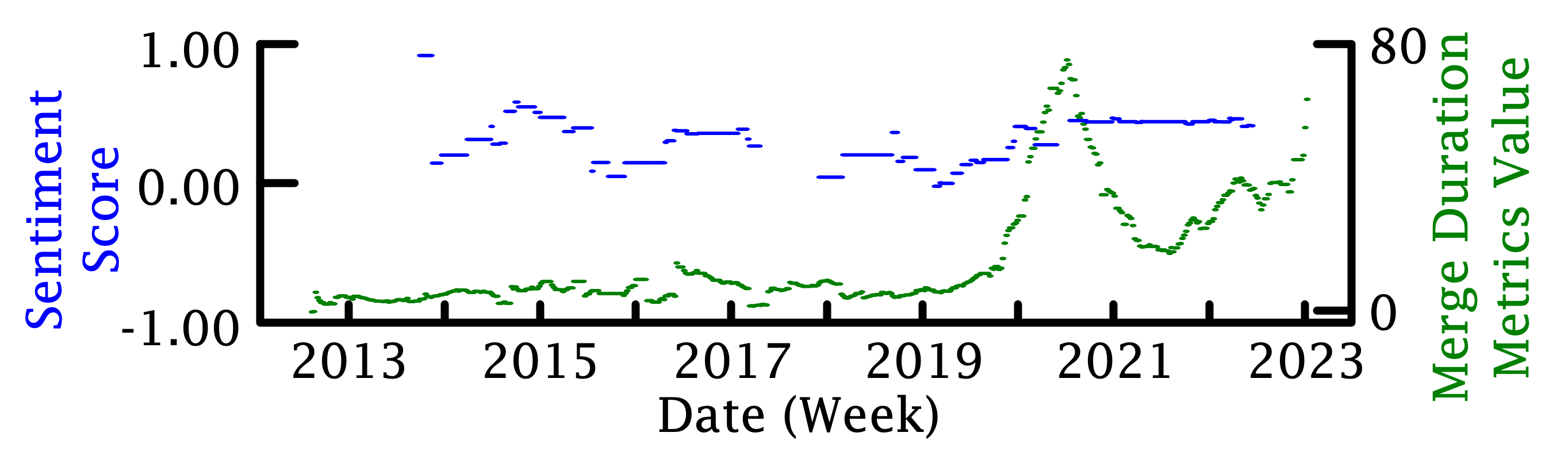}
  \caption{Time-Series Sentiment Score and Merge Duration Metrics Value of the Paintroid App}\label{fig4}
\end{figure}

\begin{figure}[tb]
  \centering
  \includegraphics[width=\linewidth]{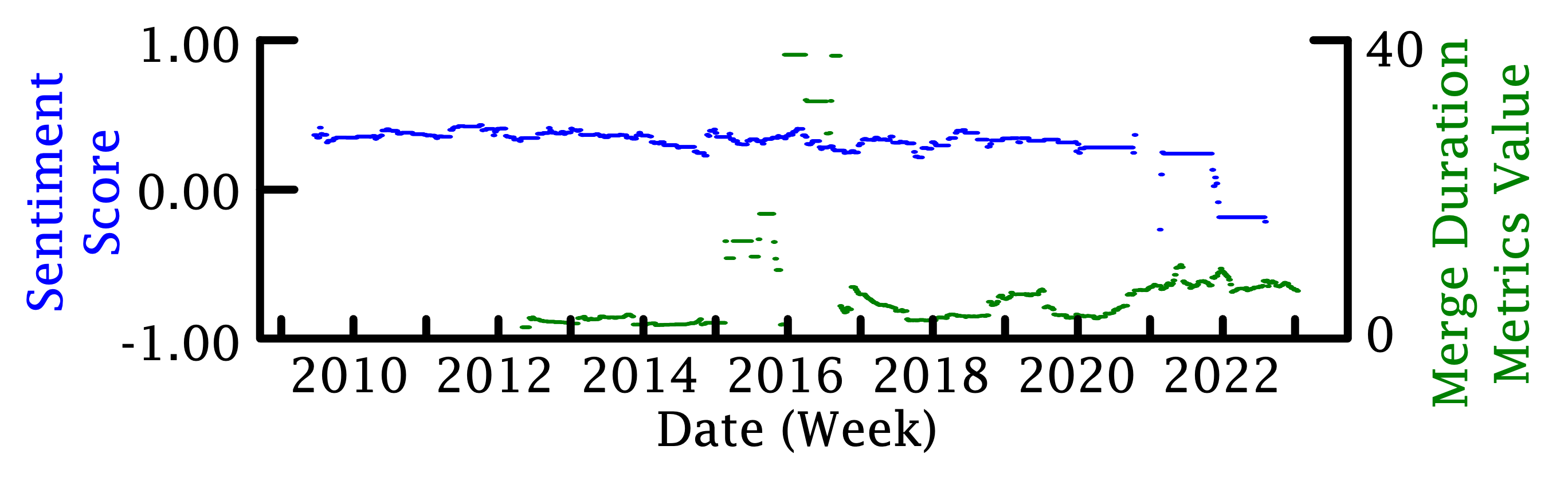}
  \caption{Time-Series Sentiment Score and Merge Duration Metrics Value of the AnySoftKeyboard App}\label{fig5}
\end{figure}

\begin{figure}[tb]
  \centering
  \includegraphics[width=\linewidth]{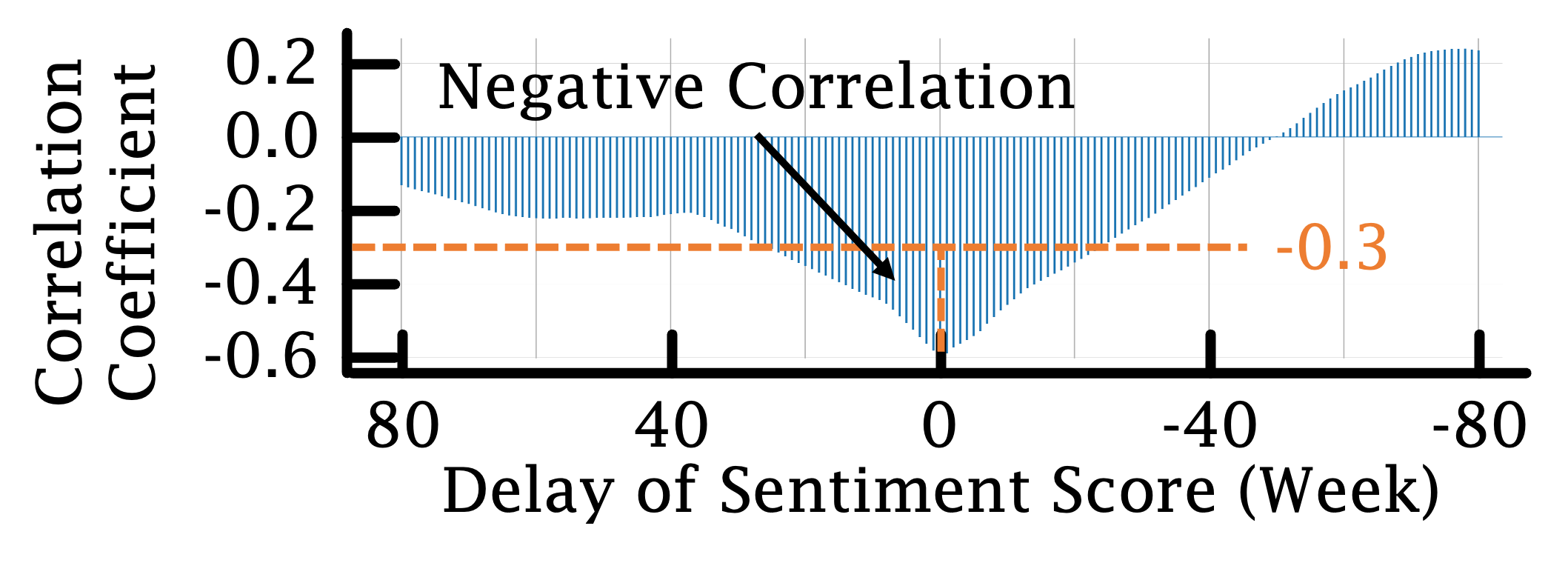}
  \caption{Correlation Coefficient Between Sentiment Score and Merge Duration Metrics Value for Each Delay of the Wire App}\label{fig6}
\end{figure}

\begin{figure}[tb]
  \centering
  \includegraphics[width=\linewidth]{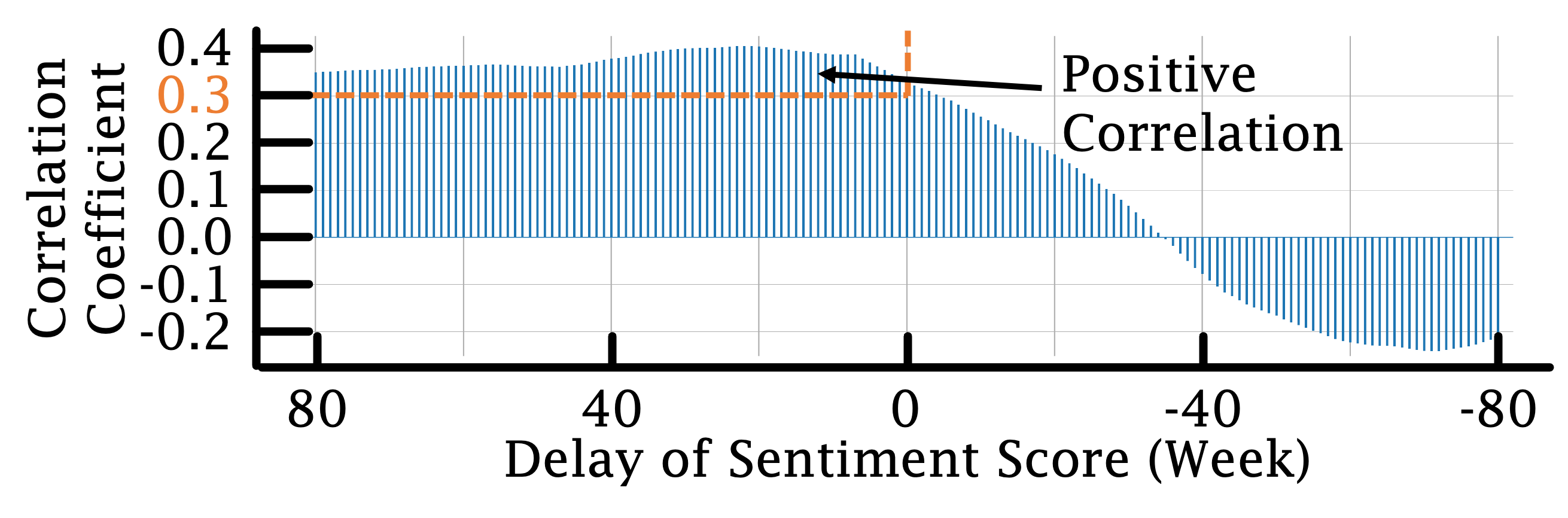}
  \caption{Correlation Coefficient Between Sentiment Score and Merge Duration Metrics Value for Each Delay of the Paintroid App}\label{fig7}
\end{figure}

\begin{figure}[tb]
  \centering
  \includegraphics[width=\linewidth]{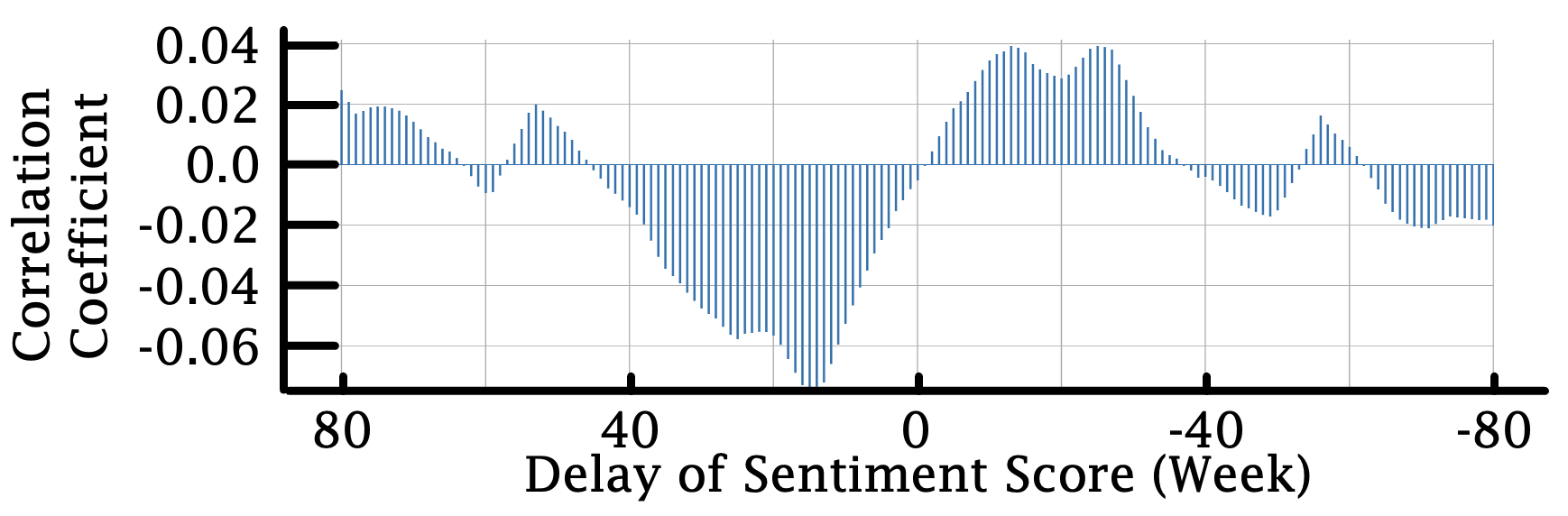}
  \caption{Correlation Coefficient Between Sentiment Score and Merge Duration Metrics Value for Each Delay of the AnySoftKeyboard App}\label{fig8}
\end{figure}

For example, Figure \ref{fig3} shows the transition between the sentiment score of user reviews and the Merge Duration metric score of the messaging app Wire. The correlation coefficient was –0.3 or less within an 80-week delay (Figure \ref{fig6}), indicating that the Merge Duration metric score for the Wire app had a negative correlation with the sentiment score of user reviews. Therefore, we considered that the correlation between the sentiment score of the user review and the Merge Duration metric score for the Wire app was negative. In the same way, Figures \ref{fig4} and \ref{fig7} show a positive correlation, whereas Figures \ref{fig5} and \ref{fig8} indicate no correlation between the metric and sentiment scores.

\begin{table}
    \caption{Percentage by Correlation between Each Development Metric and User Review Sentiment Score for All Apps (N=35) (Numbers in Brackets for Low Performing Apps (N=14))}
  \label{tab:table5}
    \begin{tabular}{cccc}
    \toprule
         Metric & Positive & Negative & No Correlations \\
    \midrule
        M1 & 26\% (21\%) & 43\% (43\%) & 31\% (36\%) \\
        M2 & 31\% (43\%) & 34\% (36\%) & 35\% (21\%) \\
        M3 & 20\% (14\%) & 60\% (64\%) & 20\% (22\%) \\
        M4 & 26\% (7\%) & 40\% (64\%) & 34\% (29\%) \\
        M5 & 17\% (7\%) & 63\% (71\%) & 20\% (22\%) \\
    \bottomrule
    \end{tabular}
\end{table}

\begin{table*}
    \caption{List of Apps Targeted for the Measurements}
  \label{tab:table7}
    \begin{tabular}{ccccc}
    \toprule
         Category & Name & \begin{tabular}{c}Interval between \\First and Last Release \\(in Years)\end{tabular}  & Number of English Text Reviews & Number of Contributors \\
    \midrule
        File Manager & Amaze & 8.1 & 1711 & 127 \\
        Flashcard & Anki & 12.1 & 11232 & 344 \\
        Podcast & AntennaPod & 10.2 & 4943 & 229 \\
        Browser & Firefox & 12.0 & 82361 & 269 \\
        Browser & DuckDuckGo & 11.8 & 369525 & 76 \\
        Messaging & Conversations & 8.8 & 348 & 128\\
        Messaging & Element & 3.7 & 501 & 375\\
        Messaging & Session & 2.9 & 732 & 100\\
        Messaging  & JitsiMeet & 6.0 & 2008 & 380\\
        Messaging  & Wire & 7.7 & 3772 & 53\\
        Note & Joplin & 4.9 & 956 & 513 \\
        Note & Markor & 5.3 & 1076 & 107 \\
        Community Clients & Slide & 5.9 & 1770 & 98 \\
        Community Clients & Tusky & 5.7 & 876 & 207 \\
        Utilities & PDFConverter & 4.5 & 351 & 142\\
        Keyboard & AnySoftKeyboard & 2.5+ & 3545 & 121\\
        Launcher \& Desktop & KISS & 7.6 & 397 & 245\\
        Password \& Authentication & Aegis & 3.8 & 561 & 47 \\
        Password \& Authentication & Bitwarden & 6.4 & 7910 & 106 \\
        Password \& Authentication & OpenKeychain & 10.0 & 668 & 109 \\
        Synchronisation & Syncthing & 8.7 & 963 & 52 \\
        Synchronisation & ownCloud & 10.4 & 870 & 87 \\
        Synchronisation & Nextcloud & 6.5 & 1102 & 174\\
        Maps \& Navigation & OsmAnd & 12.4 & 6100 & 858 \\
        VPN & MysteriumVPN & 4.0 & 832 & 24\\
        Media Frontends & Jellyfin & 2.8 & 594 & 264\\
        Science \& Education & Wikipedia & 10.9 & 32972 & 107 \\
        Automation & HomeAssistant & 3.1 & 701 & 163\\
        Mail & K-9 & 2.5+ & 5920 & 241 \\
        Mail & Tutanota & 8.0 & 1478 & 66 \\
        Painting & PocketPaint & 8.8 & 2080 & 67 \\
        Calendar & SimpleCalendar & 4.2 & 2353 & 185 \\
        Video Player & Kodi & 7.6 & 28594 & 825 \\
        Miscellaneous & Wikipedia & 12.9 & 19256 & 179 \\
        Miscellaneous &Noice & 3.5 & 157 & 131 \\
    \bottomrule
    \end{tabular}
\end{table*}

Two of the 37 apps were excluded due to implementation reasons at the time of data collection. Numbers outside brackets in Table \ref{tab:table5} show the results for the remaining 35 apps (Table \ref{tab:table7}), while numbers in brackets in Table \ref{tab:table5} show the results of the 14 apps that exhibited a general downward trend in the sentiment score of the user reviews over time. For M3 and M5, 60–63\% of all apps showed a negative correlation with the sentiment score over time (numbers outside brackets in Table \ref{tab:table5}). Similarly, 64–71\% of apps that showed a general downward trend in the sentiment score of user reviews over time had a negative correlation with the sentiment score for M3, M4, and M5 (numbers in brackets in Table \ref{tab:table5}).

\section{Discussion}

\subsection{RQ1. Can the characteristics related to the Agile development process be measured and evaluated?}

To measure and evaluate the characteristics related to the Agile development process, this study identified some development metrics relevant to Agile development, which are shown in Table \ref{tab:table2}. Through the changes in time-series trends of these metrics, some of the characteristics can be measured and evaluated.

\smallbreak

\noindent \fbox { \parbox { \linewidth} {RQ1. \textbf{Some development metrics are defined to measure and evaluate the characteristics related to the Agile development process through the changes in time-series trends of these metrics.}}}

\subsection{RQ2. Is there a relationship between the characteristics related to the Agile development process and user satisfaction?}

This study analyzed the time-series correlations between the development metrics identified in RQ1 and the sentiment scores of user text reviews, which represented user satisfaction. Regardless of the trend of the sentiment scores of the user reviews of apps over time, M3 and M5 show negative correlations with the changes in user satisfaction for most apps. Thus, there is a relationship between the characteristics related to the Agile development process and user satisfaction.

\smallbreak

\noindent \fbox { \parbox { \linewidth} {RQ2. \textbf{There is relationship since M3 (Merge Duration) and M5 (Remaining Pull Requests Lifetime) defined in RQ1 show negative correlations with changes in user satisfaction for most apps.}}}

\subsection{RQ3. What approaches can be applied in the development process to improve user satisfaction?}

This study analyzed development metrics with significant as well as weak relationships based on the time-series correlations analyzed in RQ2 with factors influencing metric changes.

Regardless of the trend of sentiment scores of the user reviews of apps over time, M3 and M5 show negative correlations with the changes in user satisfaction for most apps, indicating that a shorter time between the creation of a Pull Request and its merge is correlated with a better sentiment score for user reviews. The time between the creation of a Pull Request and its merge is considered to be directly affected by the approval speed of the work and the code conflicts or workflow rules between approval and merge. The approval speed is strongly influenced by the motivation of the people involved, the complexity of the work unit, and the appropriateness of the approval rules, while code conflicts are strongly influenced by the appropriateness of the work unit and the maintainability of the code base. Thus, the changes in user review sentiment scores are correlated with the development team motivation, the ability to set up appropriate work units, the adequacy of work rules, and the code maintainability. Efforts to improve these factors may lead to increased user satisfaction. 

For apps that show a generally downward trend in user reviews over time, development metrics with correlations greater than 60\% (M3, M4 and M5) show increases in the percentage of negative correlations ranging from 4\% to 24\%, and decreases in the percentage of positive correlations ranging from 19\% to 6\% compared to the results for all apps. This suggests that efforts to improve the factors indicated are not sufficient conditions to improve user satisfaction, but are worth considering as necessary conditions. 

In all cases, M1 and M2 do not show a strong correlation with changes in the sentiment scores, suggesting that changes in the release frequency or workload are uncorrelated with user satisfaction. Hence, the frequency of releases or workload are not factors. 

\smallbreak

\noindent \fbox { \parbox { \linewidth} {RQ3. \textbf{Approaches to improve team motivation, the ability to set up appropriate work units, the adequacy of work rules, and the code maintainability are considered to be not sufficient but necessary. Approaches to improve release frequency and workload are considered to have little effect.}}}

\subsection{Limitation and Threats to Validity}

This study is conducted within various limitations. Those limitations may introduce threats to validity. By including only OSS Android apps from the Google Play Store and using their GitHub data and user reviews, for example, the common characteristic of the development team that they use GitHub may introduce bias. In addition, observable user reviews do not always represent overall user satisfaction. Moreover, not only is there a gap between observable user reviews and true user satisfaction, but there may also be a varying degree of gap between different apps depending on the characteristics of the app and the time period and so on. Furthermore, although this study focused on apps that showed characteristics of Agile, they are not exactly developed in an Agile method. Finally, the method of deciding the moving window and calculating the metrics may not be the most reasonable one.

\section{Conclusion}

This study considered a method to examine the time-series correlations in Agile development using data on user satisfaction and development process characteristics. Some development metrics are correlated with the sentiment change of user reviews, suggesting that the motivation of the development team, the ability to set up appropriate work units, the appropriateness of development rules, and the improvement of code maintainability are insufficient conditions for improving user satisfaction, but they are worth considering as necessary ones. The results also suggest that changes in the release frequency and workload are uncorrelated with user satisfaction.

For researchers, this study suggested that a similar approach could be used to identify some more specific ominous smells in Agile development. For developers, specific ominous smells are presented and it will be possible to take care of them during development.

In the future, we plan to reconsider the selection criteria for projects to be analyzed, add development metrics, conduct qualitative examinations, improve the method for calculating metrics, and refine the analysis methodology. Projects that are not suitable for analysis based on the trend of sentiment scores of user reviews should be excluded. Additionally, different metrics should be considered. Improvements to the metrics calculation method include considering the importance of Pull Requests and Issues, while analysis refinements include considering alternate methods such as multiple regression analysis.

\bibliographystyle{ACM-Reference-Format}
\bibliography{references}


\end{document}